# Angular dependence of $\eta$ photoproduction in photon-induced reaction

Jun-Zhen Wang and Bao-Chun Li

*Department of Physics, Shanxi University, Taiyuan, Shanxi 030006, China*

Correspondence should be addressed to Bao-Chun Li; libc2010@163.com

**Abstract:** Photoproduction of $\eta$ mesons from nucleons can provide valuable information about the excitation spectrum of the nucleons. The angular dependence of $\eta$ photoproduction in the photon-induced reaction is investigated in the multi-source thermal model. The results are compared with experimental data from the $\eta \to 3\pi^0 \to 6\gamma$ decay mode. They are in good agreement with the experimental data. It is shown that the movement factor increases linearly with the photon beam energies. And, the deformation and translation of emission sources are visually given in the formalism.

## 1. Introduction

The excitation spectrum of nucleons is important to understand the non-perturbative behavior of the fundamental theory of strong interactions, Quantum Chromodynamics (QCD) [1-4]. The photon-induced meson production off nucleons mainly is used to achieve more information from the excitation spectrum of nucleons. It is an important part of the search for missing resonances that the production of $\eta$ mesons in photon-induced and hadron-induced reactions on free and quasi-free nucleons and on nuclei [5-8]. The advantage of photon-induced reactions is that the electromagnetic couplings can provide valuable information related to the details of the model wave functions. Because the electromagnetic excitations are isospin dependent, we need perform meson-production reactions off the neutron.

Recently, the photoproduction of $\eta$ mesons from quasifree protons and neutrons was measured in $\eta \to 3\pi^0 \to 6\gamma$ decay mode by the CBELSA/TAPS detector at the electron accelerator ELSA in Bonn [9]. At different incident photon energies, the experiments are performed by the incident photon beam on a liquid deuterium target. A great number of $\eta$ mesons are produced in the photon-induced reaction. The experimental data are regarded as a multiparticle system. And, their angular distributions represent an obvious regularity at different incident photon energies. In order to explain the abundant experimental results, some statistical methods are proposed and developed [10-16]. In this work, we will extend a multi-source thermal model to the statistical investigation of the angular distributions in the photon-induced reaction and try to understand the $\eta$ photoproduction in the reaction. In our previous wok [17-21], the model was focused on the investigation of the particle production in intermediate-energy and high-energy collisions.

## 2. $\eta$ meson distribution in the multi-source thermal model

In the multi-source thermal model [17-21], many emission sources are expected to be formed at the final stage of the photon-induced reaction. Every source emits particles isotropically in the source rest frame. The observed $\eta$ mesons are from different emission sources. The incident beam direction is defined as an $oz$ axis and the reaction plane is defined as $yoz$ plane. In the source rest frame, the meson momentum $p_x$, $p_y$ and $p_z$ obey a normal distribution. The corresponding transverse momentum $p_T = \sqrt{p_x^2 + p_y^2}$ obeys a Rayleigh distribution

$$f_{p_T}(p_T) = \frac{p_T}{\sigma^2} e^{-\frac{p_T^2}{2\sigma^2}}, \qquad (1)$$

where $\sigma$ represents a distribution width. The distribution function of the polar angle $\theta$ is

$$f_\theta(\theta) = \frac{1}{2}\sin\theta \qquad (2)$$

Because of the interactions with other emission sources, the considered source deforms and translates along the $oz$ axis. Then, the momentum component is revised to

$$p_z' = a_z p_z + b_z \qquad (3)$$

where $a_z$ and $b_z$ represent the coefficients of the source deformation and translation along the $oz$ axis, respectively. The mathematical description of the deformable translational source is formulized simply as a linear relationship between $p_z'$ and $p_z$, which reflects the mean result of the source interaction. For $a_z \neq 1$ or $b_z \neq 0$, the $p_z'$ distribution of $\eta$ mesons is anisotropic along the $oz$ axis.

By using Monte Carlo method, the $p_T$ and $p_z'$ are given by

$$p_T = \sigma\sqrt{-2\ln r_1}, \tag{4}$$

$$p_z' = a_z \sigma\sqrt{-2\ln r_2}\cos(2\pi r_3) + b_z, \tag{5}$$

where $r_1$, $r_2$ and $r_3$ are random numbers from 0 to 1. The polar angle $\theta$ is revised to

$$\theta' = \arctan\frac{p_T}{p_z'} = \arctan\frac{\sqrt{-2\ln r_1}}{a_z\sqrt{-2\ln r_2}\cos(2\pi r_3) + b_z}. \tag{6}$$

We can calculate a new distribution function of the polar angle by this formula.

### 3. Angular dependences of $\eta$ photoproduction in the photon-induced reaction

Figs. 1(a)-1(p) show the angular distributions of $\eta$ mesons for different bins of incident photon energy $698 \leq E_\gamma \leq 1005$ as a function of $\cos\theta_\eta$. The $\theta_\eta$ is the polar angle of $\eta$ meson in the beam-target cm system assuming the initial state nucleon at rest. The symbols represent the experimental data from the CBELSA/TAPS detector at the electron accelerator ELSA in Bonn [9]. The results obtained by using the multi-source thermal model are shown with the curves, which behave the same as the experimental data in the 16 bins of incident photon energy. By minimizing $\chi^2$ per degree of freedom ($\chi^2/\text{dof}$), we determine the corresponding parameters $a_z$ and $b_z$, which are presented in Table 1. It is found that there is an almost linear relationship between the $b_z$ and $E_\gamma$. As representative energies of Figs. 1, we give a schematic sketch of these emission sources at the four different energies in Figs. 7 (a). The deformations and translations can be seen intuitively in the figure.

In Figs. 2(a)-2(p) and Figs. 3(a)-3(p), we present the angular distributions of $\eta$ mesons for different bins of incident photon energy $1035 \leq E_\gamma \leq 1835$ as a function of $\cos\theta_\eta$. The $\theta_\eta$ is the polar angle of $\eta$ meson in the beam-target cm system assuming the initial state nucleon at rest. Same as Figs. 1, the symbols represent the experimental data from the CBELSA/TAPS detector at the electron accelerator ELSA in Bonn [9]. The results obtained by using the multi-source thermal model are shown with the curves, which behave the same as the experimental data in the 28 bins of incident photon energy. The parameters $a_z$ and $b_z$ are presented in Table 2 and Table 3. As the representative energies of Figs. 1 and Figs. 2, the schematic sketches of the emission sources are given at different energies in Figs. 7 (b) and Figs. 7 (c).

In Figs. 4, Figs. 5 and Figs. 6, we show angular distributions in the $\eta$-nucleon cm system for $\gamma p \to p\eta$ reaction for the different bins of final state energy $1488 \leq W \leq 1625$, $1635 \leq W \leq 1830$ and $1850 \leq W \leq 2070$, respectively. Same as Figs. 1, the model results and experimental data are indicated by the curves and symbols, respectively. The model results can also agree with the experimental data. In the same way, the deformations and translations of these emission sources are given in Table 4-6 and Figs. 7 (d)-7 (f). All the parameter values taken in the above calculations are also given in Figs. 8 and Figs. 9. It can be found that the $a_z$ keeps almost invariable and fluctuates around 1.0 with the increasing $E_\gamma$. The $b_z$ increases linearly with the increasing $E_\gamma$, i.e., $b_z = (0.541 \pm 0.005) \times 10^{-3} E_\gamma - (0.622 \pm 0.011)$. There are similar relationships between the parameters and different final state energies $W$, where the fitting function of $b_z$ is $b_z = (0.808 \pm 0.003) \times 10^{-3} W - (1.322 \pm 0.007)$.

**4. Discussion and Conclusions**

The excitation spectrum of nucleons can especially help us to understand the strong interaction in the non-perturbative regime. Before, the hadron induced reactions is a main experimental method in the investigation. In the last two decades, the photon-induced reaction and electron scattering experiment are applied to study the electromagnetic excitation of baryons. Recently, the photoproduction of $\eta$ mesons from quasifree protons and neutrons are measured by the CBELSA/TAPS detector. In the paper, we theoretically study the angular distribution of $\eta$ mesons for different incident photon energies $E_\gamma$ and for different final state energies $W$. Then, the results are compared with the experimental data in detail. The deformation coefficient $a_z$ and translation coefficient $b_z$ are extracted by the comparison. The $a_z$ is almost independent of incident photon energies and final state energies. The $b_z$ is linearly dependent on incident photon energies and final state energies. In particular, we visually give the deformation and translation of the emission sources by schematic sketches. From the patterns, it is intuitive and easy to better understand the motion and configuration of the emission sources.

A great number of $\eta$ mesons are produced in the photon-induced reaction. These $\eta$ mesons are regarded as a multiparticle system, which can be analyzed by the statistical method. In recent years, we develop such a model, which is called multi-source thermal model. Some emission sources of final-state particles are formed in the reaction. Each emission source emits particles isotropically in the rest frame of the emission source. Due to the source interaction, the sources emit particles anisotropically. The $\eta$ mesons are emitted from these sources. In our previous work, the model can successfully describe transverse momentum spectra and pseudorapidity spectra of final-state particles produced in proton-proton ($pp$) collisions, proton-nucleus ($pA$) collisions and nucleus-nucleus ($AA$) collisions at intermediate energy and at high energy [17-21]. In this work, we extend the multi-source thermal model to the statistical investigation of final-state particles produced in the photon-induced reaction. The model is improved to describe the angular dependence of the $\eta$ photoproduction from quasifree protons and neutrons. The information of the source deformation and translation is obtained with different beam energies. It is helpful for us to understand the $\eta$ photoproduction.

**Acknowledgments**

This work is supported by the National Natural Science Foundation of China under Grants No. 11247250 and No. 11575103, the Shanxi Provincial Natural Science Foundation under Grants No. 201701D121005.

**References**

[1] B. Krusche and C. Wilkin, Prog. Part. Nucl. Phys. **80**, 43 (2014).

[2] M.~Dieterle *it et al.*, Phys. Lett. B **770**, 523 (2017).

[3] E. Y. Paryev, J. Phys. G **40**, 025201 (2013).

[4] T. Sekihara, H. Fujioka and T. Ishikawa, Phys. Rev. C **97**, 045202 (2018).

[5] A. V. Anisovich *et al.*, Phys. Rev. C **96**, 055202 (2017).

[6] V. Kuznetsov *et al.*, JETP Lett. **106**, 693 (2017).

[7] D. Werthmüller *et al.* [A2 Collaboration], Phys. Rev. C **90**, 015205 (2014).

[8] B. Krusche, Prog. Part. Nucl. Phys. **67**, 412 (2012).

[9] L. Witthauer *et al.* [CBELSA/TAPS Collaboration], Eur. Phys. J. A **53**, 58 (2017).

[10] J. Rafelski and J. Letessier, Nuclear Physics A **715**, 2003.

[11] A. Andronic, P. Braun-Munzinger and J. Stachel, Phys. Lett. B **673**, 142 (2009).

[12] J. Cleymans *et al.*, Phys. Rev. C **74**, 034903 (2006).

[13] P. Braun-Munzinger, J. Stachel and C. Wetterich, Phys. Lett. B **596**, 61 (2004).

[14] A. Adare *et al.* (PHENIX Collaboration), Phys. Rev. D **83**, 052004 (2011).

[15] K. Aamodt *et al.* (ALICE Collaboration), Phys. Lett. B **693**, 53 (2010).

[16] C. Y. Wong and G. Wilk, Phys. Rev. D **87**, 114007 (2013).

[17] B. C. Li, Y. Y. Fu, L. L. Wang, E. Q. Wang, and F. H. Liu, J. Phys. G **39**, 025009 (2012).

[18] F. H. Liu, L. N. Gao and R. A. Lacey, Adv. High Energy Phys. **2016**, 9467194 (2016).

[19] B. C. Li, Y. Z. Wang, F. H. Liu, X. J. Wen and Y. E. Dong, Phys. Rev. D **89**, 054014 (2014).

[20] B. C. Li, Y. Z. Wang and F. H. Liu, Phys. Lett. B **725**, 352 (2013).

[21] B. C. Li, Z. Zhang, J. H. Kang, G. X. Zhang and F. H. Liu, Adv. High Energy Phys. **2015**, 741816 (2015).

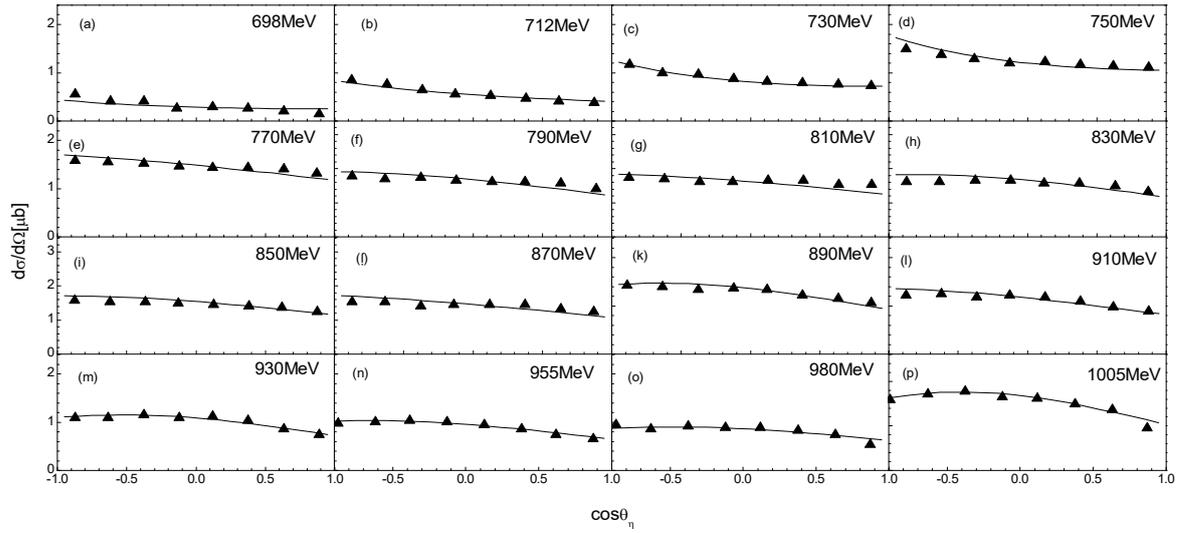

FIG. 1. Angular distributions for different bins of incident photon energy $698 \leq E_\gamma \leq 1005$ as a function of $\cos\theta_\eta$ in the beam-target cm system assuming the initial state nucleon at rest. The symbols represent the experimental data from the CBELSA/TAPS detector at the electron accelerator ELSA in Bonn [9]. The results in the multi-source thermal model are shown with the curves.

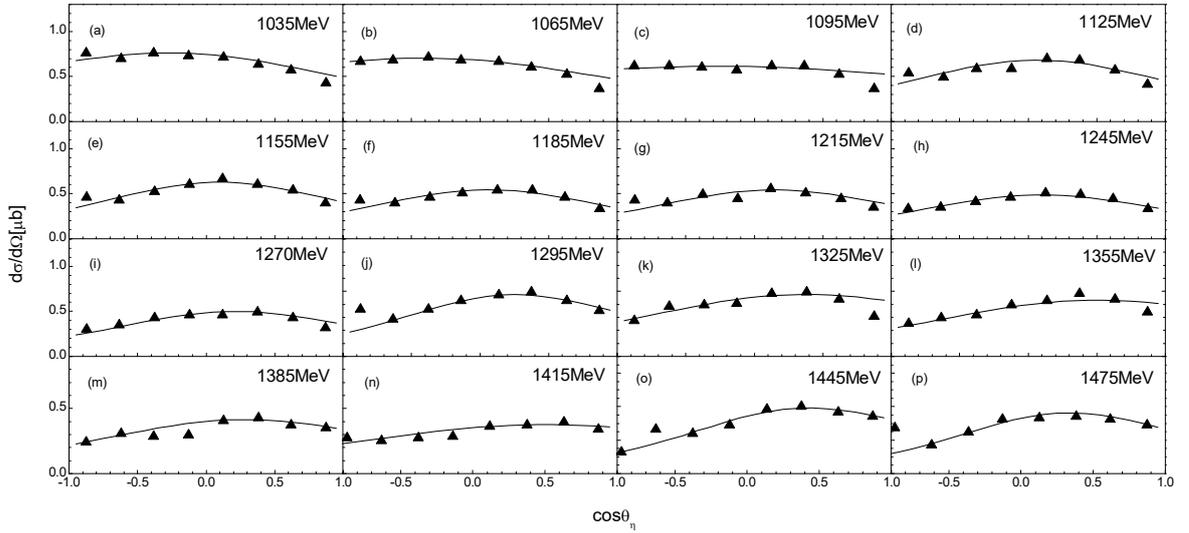

FIG. 2. Same as Figs. 1, but showing angular distributions for different bins of incident photon energy $1035 \leq E_\gamma \leq 1475$.

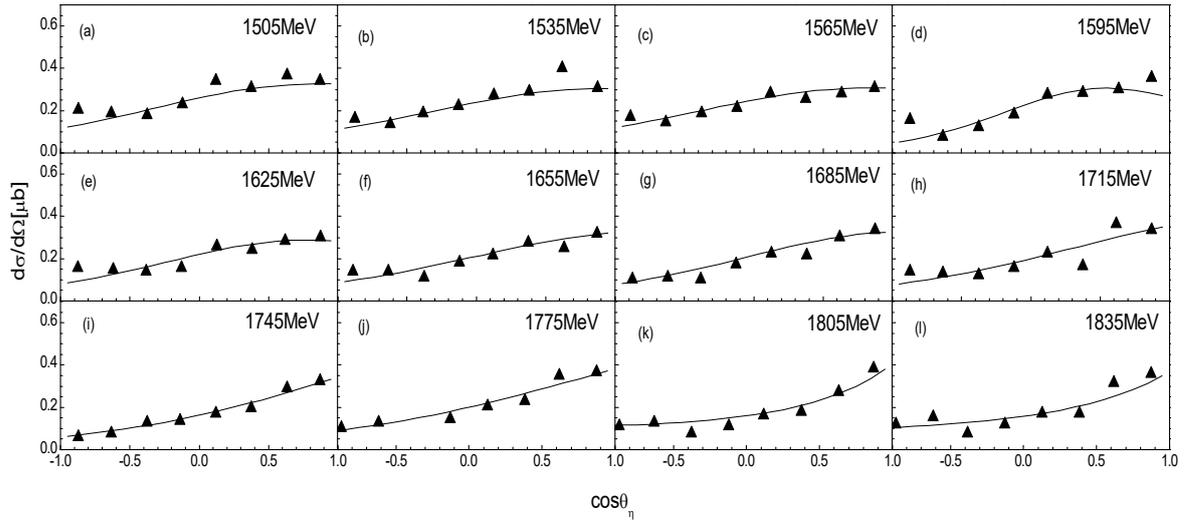

FIG. 3. Same as Figs. 1, but showing angular distributions for different bins of incident photon energy $1505 \leq E_\gamma \leq 1835$.

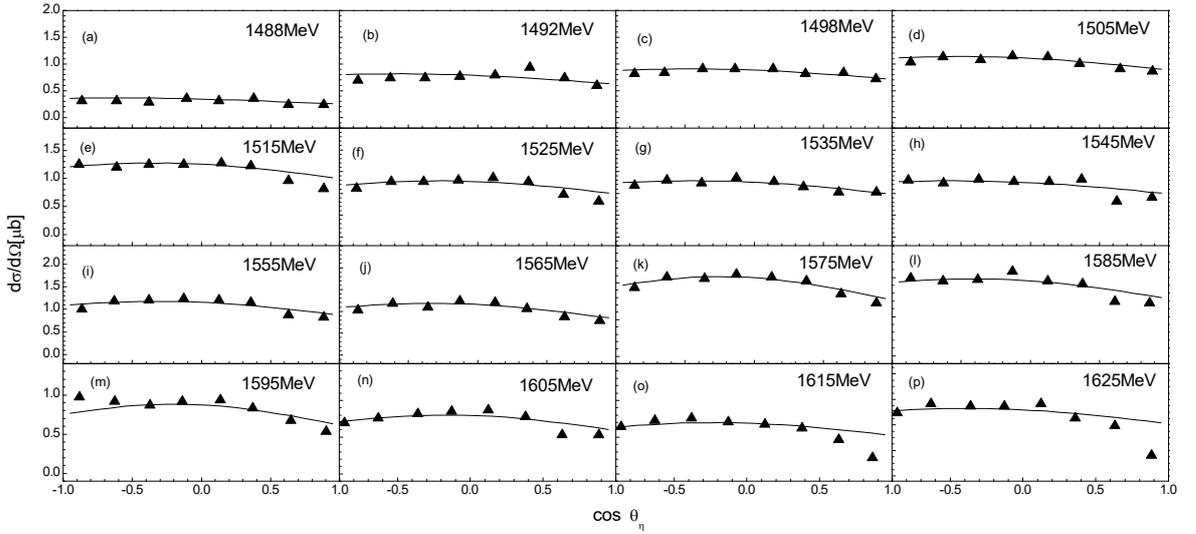

FIG. 4. Angular distributions in the $\eta$-nucleon cm system for the reaction $\gamma p \to p\eta$ for the different bins of final state energy $1488 \leq W \leq 1625$. The symbols represent the experimental data from the CBELSA/TAPS detector at the electron accelerator ELSA in Bonn [9]. The results in the multi-source thermal model are shown with the curves.

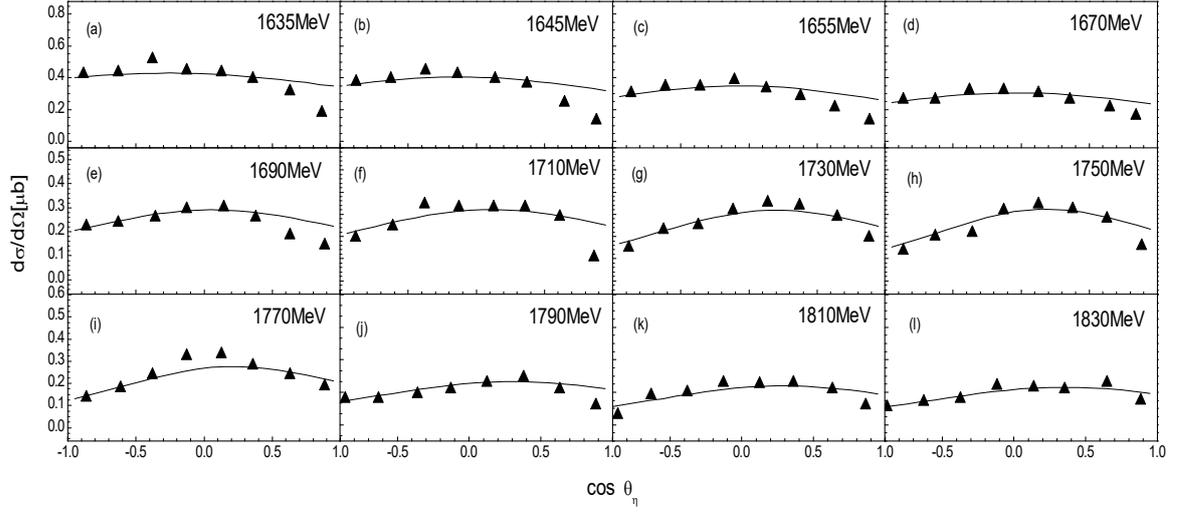

FIG. 5. Same as Figs. 4, but showing angular distributions for the different bins of final state energy $1635 \leq W \leq 1830$.

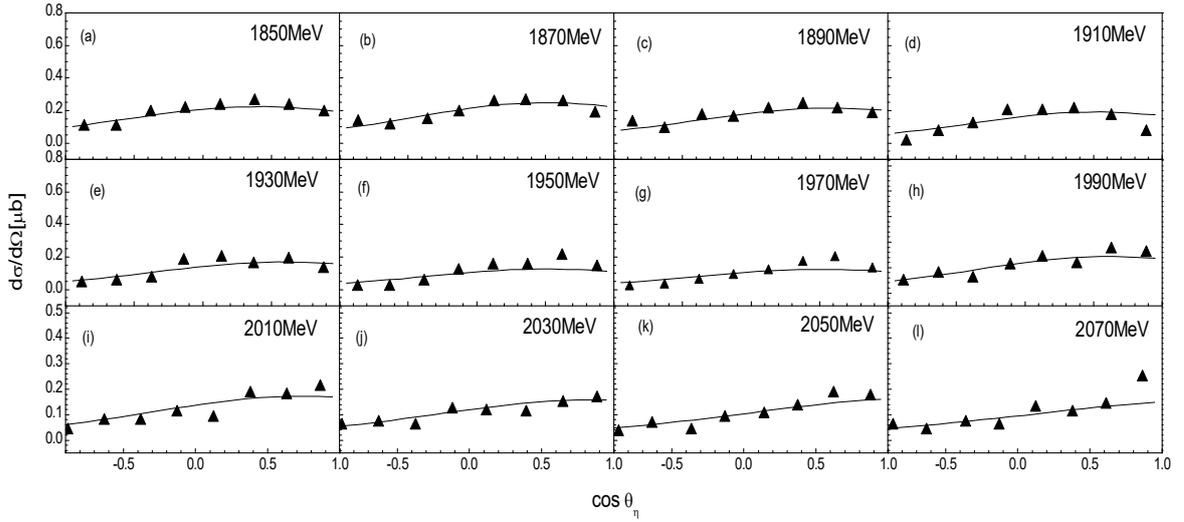

FIG. 6. Same as Figs. 4, but showing angular distributions for the different bins of final state energy $1850 \leq W \leq 2070$.

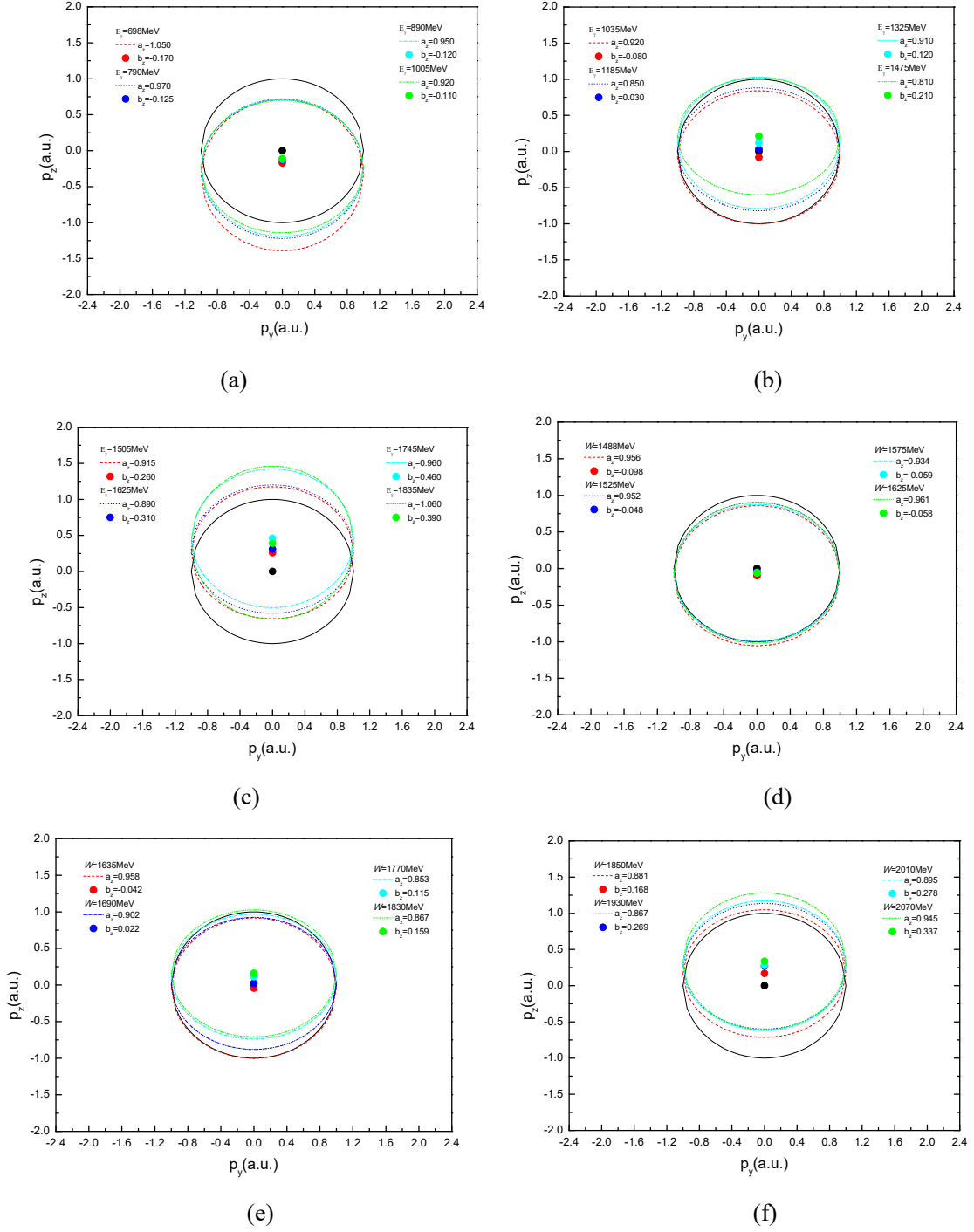

FIG. 7. The deformable and translational source in the reaction plane for different bins of incident photon energy $E_\gamma$ or final state energy $W$.

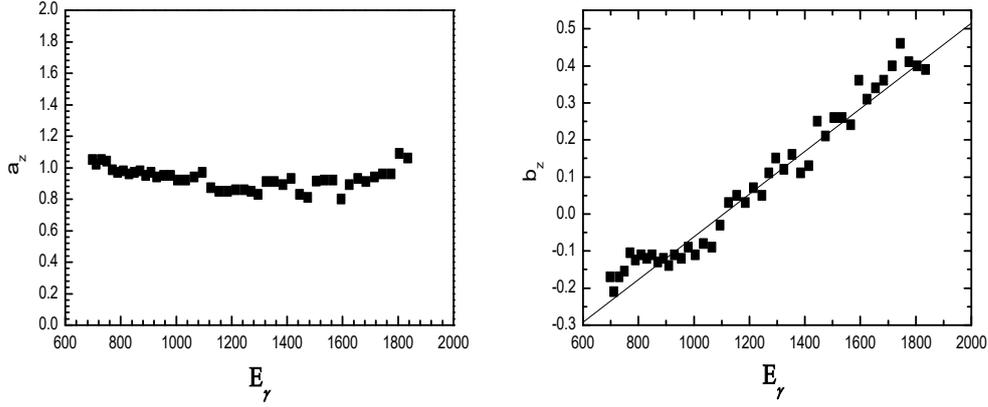

FIG. 8. The $a_z$ and $b_z$ for different bins of incident photon energy $E_\gamma$. The symbols are the values taken in Figs.1-3. The straight line is a fitted curve.

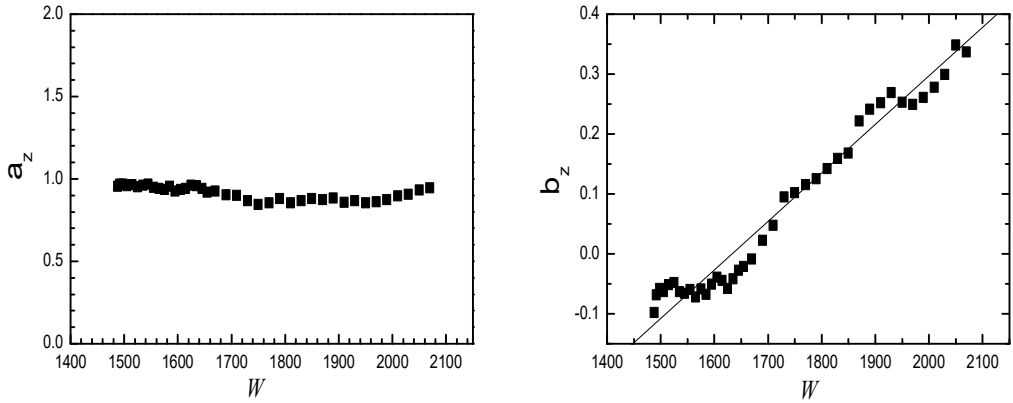

FIG. 9. The $a_z$ and $b_z$ values taken for different bins of final state energy $W$. The symbols are the values taken in Figs. 4-6. The straight line is a fitted curve.

TABLE 1.  Values of $a_z$ and $b_z$ taken in Figs. 1 model results.

| Figs. 1 | $E_\gamma$ (MeV) | $a_z$ | $b_z$ | $\chi^2/dof$ |
|---|---|---|---|---|
| (a) | 698 | 1.050 | -0.170 | 0.118 |
| (b) | 712 | 1.020 | -0.210 | 0.105 |
| (c) | 730 | 1.050 | -0.170 | 0.090 |
| (d) | 750 | 1.040 | -0.155 | 0.134 |
| (e) | 770 | 0.985 | -0.105 | 0.182 |
| (f) | 790 | 0.970 | -0.125 | 0.179 |
| (g) | 810 | 0.979 | -0.110 | 0.194 |
| (h) | 830 | 0.960 | -0.120 | 0.192 |
| (i) | 850 | 0.970 | -0.110 | 0.200 |
| (j) | 870 | 0.980 | -0.130 | 0.211 |
| (k) | 890 | 0.950 | -0.120 | 0.175 |
| (l) | 910 | 0.970 | -0.140 | 0.261 |
| (m) | 930 | 0.940 | -0.110 | 0.179 |
| (n) | 955 | 0.950 | -0.120 | 0.154 |
| (o) | 980 | 0.950 | -0.090 | 0.138 |
| (p) | 1005 | 0.920 | -0.110 | 0.150 |

TABLE 2. Values of $a_z$ and $b_z$ taken in Figs. 2 model results.

| Figs. 2 | $E_\gamma$ (MeV) | $a_z$ | $b_z$ | $\chi^2/dof$ |
|---|---|---|---|---|
| (a) | 1035 | 0.920 | -0.080 | 0.124 |
| (b) | 1065 | 0.940 | -0.090 | 0.118 |
| (c) | 1095 | 0.970 | -0.030 | 0.132 |
| (d) | 1125 | 0.870 | 0.030 | 0.165 |
| (e) | 1155 | 0.850 | 0.050 | 0.170 |
| (f) | 1185 | 0.850 | 0.030 | 0.168 |
| (g) | 1215 | 0.860 | 0.070 | 0.173 |
| (h) | 1245 | 0.860 | 0.050 | 0.122 |
| (i) | 1270 | 0.850 | 0.110 | 0.121 |
| (j) | 1295 | 0.830 | 0.150 | 0.145 |
| (k) | 1325 | 0.910 | 0.120 | 0.147 |
| (l) | 1355 | 0.910 | 0.160 | 0.142 |
| (m) | 1385 | 0.890 | 0.110 | 0.115 |
| (n) | 1415 | 0.930 | 0.130 | 0.106 |
| (o) | 1445 | 0.830 | 0.250 | 0.122 |
| (p) | 1475 | 0.810 | 0.210 | 0.091 |

TABLE 3. Values of $a_z$ and $b_z$ taken in Figs. 3 model results.

| Figs. 3 | $E_\gamma$ ( MeV) | $a_z$ | $b_z$ | $\chi^2/dof$ |
|---|---|---|---|---|
| (a) | 1505 | 0.915 | 0.260 | 0.243 |
| (b) | 1535 | 0.920 | 0.260 | 0.190 |
| (c) | 1565 | 0.920 | 0.240 | 0.158 |
| (d) | 1595 | 0.800 | 0.360 | 0.315 |
| (e) | 1625 | 0.890 | 0.310 | 0.179 |
| (f) | 1655 | 0.930 | 0.340 | 0.190 |
| (g) | 1685 | 0.910 | 0.360 | 0.206 |
| (h) | 1715 | 0.940 | 0.400 | 0.235 |
| (i) | 1745 | 0.960 | 0.460 | 0.085 |
| (j) | 1775 | 0.960 | 0.410 | 0.144 |
| (k) | 1805 | 1.090 | 0.400 | 0.132 |
| (l) | 1835 | 1.060 | 0.390 | 0.140 |

TABLE 4. Values of $a_z$ and $b_z$ taken in Figs. 4 model results.

| Figs. 4 | $W$ ( MeV) | $a_z$ | $b_z$ | $\chi^2/dof$ |
|---|---|---|---|---|
| (a) | 1488 | 0.956 | -0.098 | 0.086 |
| (b) | 1492 | 0.968 | -0.069 | 0.125 |
| (c) | 1498 | 0.966 | -0.058 | 0.101 |
| (d) | 1505 | 0.959 | -0.063 | 0.114 |
| (e) | 1515 | 0.965 | -0.052 | 0.205 |
| (f) | 1525 | 0.952 | -0.048 | 0.192 |
| (g) | 1535 | 0.960 | -0.063 | 0.143 |
| (h) | 1545 | 0.967 | -0.066 | 0.206 |
| (i) | 1555 | 0.948 | -0.060 | 0.174 |
| (j) | 1565 | 0.941 | -0.072 | 0.168 |
| (k) | 1575 | 0.934 | -0.059 | 0.170 |
| (l) | 1585 | 0.955 | -0.068 | 0.235 |
| (m) | 1595 | 0.925 | -0.051 | 0.260 |
| (n) | 1605 | 0.934 | -0.039 | 0.251 |
| (o) | 1615 | 0.942 | -0.045 | 0.307 |
| (p) | 1625 | 0.961 | -0.058 | 0.293 |

TABLE 5. Values of $a_z$ and $b_z$ taken in Figs. 5 model results.

| Figs. 5 | $W$ (MeV) | $a_z$ | $b_z$ | $\chi^2/dof$ |
|---|---|---|---|---|
| (a) | 1635 | 0.958 | -0.042 | 0.305 |
| (b) | 1645 | 0.940 | -0.028 | 0.350 |
| (c) | 1655 | 0.920 | -0.021 | 0.263 |
| (d) | 1670 | 0.925 | -0.009 | 0.187 |
| (e) | 1690 | 0.902 | 0.022 | 0.240 |
| (f) | 1710 | 0.899 | 0.047 | 0.209 |
| (g) | 1730 | 0.868 | 0.095 | 0.181 |
| (h) | 1750 | 0.843 | 0.102 | 0.194 |
| (i) | 1770 | 0.853 | 0.115 | 0.235 |
| (j) | 1790 | 0.879 | 0.125 | 0.170 |
| (k) | 1810 | 0.855 | 0.142 | 0.152 |
| (l) | 1830 | 0.867 | 0.159 | 0.138 |

TABLE 6. Values of $a_z$ and $b_z$ taken in Figs. 6 model results.

| Figs. 6 | $W$ (MeV) | $a_z$ | $b_z$ | $\chi^2/dof$ |
|---|---|---|---|---|
| (a) | 1850 | 0.881 | 0.168 | 0.127 |
| (b) | 1870 | 0.872 | 0.221 | 0.132 |
| (c) | 1890 | 0.883 | 0.241 | 0.135 |
| (d) | 1910 | 0.856 | 0.252 | 0.153 |
| (e) | 1930 | 0.867 | 0.269 | 0.218 |
| (f) | 1950 | 0.853 | 0.253 | 0.295 |
| (g) | 1970 | 0.860 | 0.249 | 0.321 |
| (h) | 1990 | 0.875 | 0.261 | 0.184 |
| (i) | 2010 | 0.895 | 0.278 | 0.359 |
| (j) | 2030 | 0.906 | 0.299 | 0.337 |
| (k) | 2050 | 0.932 | 0.348 | 0.285 |
| (l) | 2070 | 0.945 | 0.337 | 0.304 |